\documentclass[aps,prl,twocolumn,showpacs,superscriptaddress,floatfix,longbibliography]{revtex4-2}
\usepackage[utf8]{inputenc}
\usepackage{amssymb,amsmath}

\usepackage{graphicx}
\usepackage{braket}
\usepackage{xcolor}
\usepackage{physics}
\usepackage[normalem]{ulem}

\usepackage{hyperref}
\hypersetup{
    colorlinks=true,
    linkcolor=blue,
    citecolor=blue,
    filecolor=blue,
    urlcolor=blue
}

\def\Rb87{^{87}\mathrm{Rb}}
\def\Na23{^{23}\mathrm{Na}}
\def\K40{^{40}\mathrm{K}}
\def\Force{{\mathcal F}}
\def\height{{\eta}}
\def\chemicalpotential{{\varepsilon}}

\def\ex{\mathbf{e}_x}
\def\ey{\mathbf{e}_y}
\def\ez{\mathbf{e}_z}

\def\hz_um{\mathrm{Hz /\mu m}}
\def\Hz{\mathrm{Hz}}

\begin{document}

\title{The Rayleigh-Taylor instability in a binary quantum fluid}
\author{Y.~Geng}
\thanks{These two authors contributed equally}
\author{J.~Tao}
\thanks{These two authors contributed equally}
\author{M.~Zhao}
\author{S.~Mukherjee}
\affiliation{Joint Quantum Institute, University of Maryland and National Institute of Standards and Technology, College Park, Maryland, 20742, USA}

\author{S.~Eckel}
\email{stephen.eckel@nist.gov}
\affiliation{National Institute of Standards and Technology, Gaithersburg, Maryland, 20899, USA}

\author{G.~K.~Campbell}
\email{gretchen.campbell@nist.gov}
\affiliation{Joint Quantum Institute, University of Maryland and National Institute of Standards and Technology, College Park, Maryland, 20742, USA}

\author{I.~B.~Spielman}
\email{ian.spielman@nist.gov}
\affiliation{Joint Quantum Institute, University of Maryland and National Institute of Standards and Technology, College Park, Maryland, 20742, USA}

\date{\today}

\begin{abstract}

Instabilities, where initially small fluctuations seed the formation of large-scale structures, govern dynamics in a wide variety of fluid systems.
The Rayleigh-Taylor instability (RTI)~\cite{Rayleigh1892, taylor1950instability} is an iconic example and leads to the development of mushroom-shaped incursions when immiscible fluids are forced into each other.
The RTI drives structure formation throughout science and engineering including table-top oil and water mixtures; supernova explosions~\cite{abarzhi2018supernova}; and inertial confinement fusion~\cite{Graves2012}.
Despite its ubiquity, controlled laboratory RTI experiments are difficult to achieve.
Here we report the observation of the RTI in an immiscible binary superfluid~\cite{sasaki2009rti,kobyakov2011rti,saboo2023rti} consisting of a two-component Bose-Einstein condensate of $\Na23$ atoms~\cite{stamperkurn2013spinor}.
We induce the RTI at the interface between these components using a magnetic gradient to force the components together and observe the growth of mushroom-like structures.
The interface can also be stabilized, allowing us to spectroscopically measure the ``ripplon'' interface modes~\cite{takahashi2015ngmodes, watanabe2014ngbosons, Takeuchi2013}.
Lastly, we use matter-wave interferometry to transform the superfluid velocity field at the interface into a vortex chain.
These results---in agreement with our theory---highlight the similarities and differences of the RTI in classical and quantum fluids.

\end{abstract}

\maketitle

\begin{figure*}[t]
    \centering
    \includegraphics{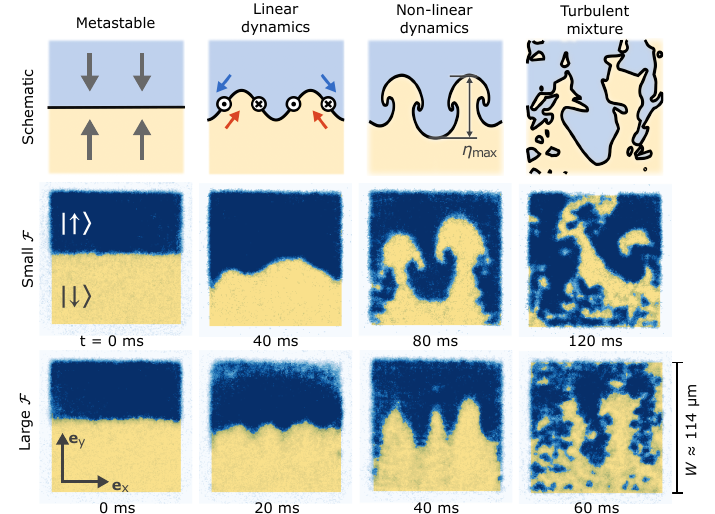}
    \caption{
    \textbf{Progression of the Rayleigh-Taylor instability showing atomic density in $\ket{\uparrow}$ (blue) and $\ket{\downarrow}$ (yellow, see Materials and Methods).}
    From top to bottom: schematic illustrations, experiments with $\Force/h = -3.1(2)\ \hz_um$ (small $\Force$), and experiments with $\Force/h = -7.7(4)\ \hz_um$ (large $\Force$).
    Initial forces (grey arrows) destabilize a fluid interface.
    The interface then develops nominally sinusoidal modulations where currents counterflowing in the two fluids (colored arrows) induce vorticity at the interface (black symbols).
    The modulations subsequently acquire characteristic mushroom- or spike-shapes before dissolving into a turbulent mixture.
    }
    \label{fig:RT_mushroom}
\end{figure*}

Fluid systems are replete with instabilities with wide-ranging impact: be it droplet formation via the Plateau–Rayleigh instability~\cite{Plateau1873,Rayleigh1892}, destabilization of fusion reactions in tokamaks via magnetohydrodynamic instabilities~\cite{Graves2012}, or galactic structure formation via fluid-gravitational instabilities~\cite{Jog1984}.
More specifically, the iconic Rayleigh-Taylor instability (RTI) ~\cite{lord1900investigation,taylor1950instability} is crucial in classical fluids across scales, from laboratory experiments to astronomical phenomena~\cite{banerjee2020classicalrti,abarzhi2018supernova,Allen1984,kilkenny1994rtifusion}.
The RTI is driven by buoyancy forces that press immiscible fluids together, such as when a fluid of higher density is placed above a lower density fluid in a gravitational potential.
Under these conditions, infinitesimal fluctuations at the horizontal interface grow exponentially: at even a minuscule local elevation increase the lighter fluid upwells, and at local depressions the denser fluid sinks.
Figure~\ref{fig:RT_mushroom} illustrates how this process unfolds: a nearly flat interface first develops sinusoidal modulations via the RTI which, in conjunction with the Kelvin-Helmholtz instability, evolve into bubble\mbox{-,} spike\mbox{-,} and mushroom-like structures that finally dissolve into a turbulent mixture.
The ubiquitous presence of the RTI in classical fluids raises natural questions: does it have analogues in quantum fluids, and, if so, how do the two relate?

\begin{figure}[t]
    \centering
    \includegraphics{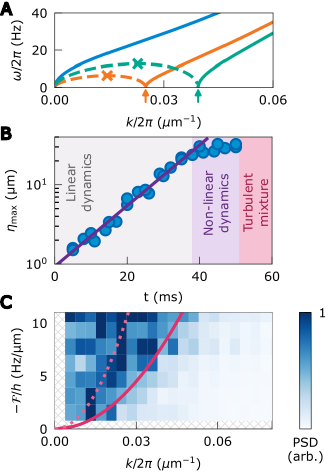}
    \caption{
    \textbf{Growth of unstable modes.}
    (\textbf{A}) Dispersion of interface excitations computed for $\Force/h = 16 \ {\rm Hz}/{\mu \rm m}$, $-3.1\ \hz_um$ and $-7.7 \ {\rm Hz}/{\mu \rm m}$ (blue, red and green respectively) using  Eq.~(\ref{eqn:dispersion}).
    Solid and dashed curves plot real and imaginary components, respectively with $k_c$ marked by arrows and $k_{\rm max}$ by crosses.
    (\textbf{B}) Interface height $\height_{\rm max}$ as a function of evolution time $t$ for $\Force/h = -7.7(4)\ \hz_um$.
    The solid curve is an exponential fit to data in the ``linear dynamics" regime.
    (\textbf{C}) Power spectrum of interface height measured during the exponential growth regime.
    The red curve plots the threshold wave vector $k_{\rm c}$ and the dashed curve plots wave vector with maximum gain $k_{\rm max}$.
    }
    \label{fig:unstable_k}
\end{figure}

Two-component Bose-Einstein condensates (BECs) with ferromagnetic interactions are ideal RTI candidates~\cite{sasaki2009rti,kobyakov2011rti,saboo2023rti}
as strong repulsion between atoms in different internal states drives phase separation~\cite{stamperkurn2013spinor} and magnetic gradient forces can rapidly switch between stable and metastable configurations.
In both configurations, interface waves with small height fluctuations $\propto \cos(k x - \omega t)$ are well described by a linearized model~\cite{landau1987fluid,kobyakov2011rti},
which predicts the dispersion relation
\begin{align}
\omega^2 &= \frac{1}{2m} \left(\Force k + \frac{\sigma}{\bar\rho} k^3\right). \label{eqn:dispersion}
\end{align}
that connects the wave vector $k$ to the angular frequency $\omega$: the same as for classical fluid interfaces.
Here, $\Force$ is the differential force between the layers; $m$ is the fluid particle mass; $\bar \rho$ is the average number density of the fluids; and $\sigma$ is the interfacial tension, which is analogous to surface tension at a liquid-gas interface.
In the stable configuration, $\Force > 0$, the frequency $\omega$ is always real (blue curve in Fig.~\ref{fig:unstable_k}A),
Consequently, initial fluctuations at the interface travel as waves known as gravity-capillary waves in classical fluids, or ripplons in superfluids.
As a type of quantum excitations, ripplons are of particular theoretical interest~\cite{takahashi2015ngmodes, watanabe2014ngbosons, Takeuchi2013} as their energy is a fractional power of momentum: neither quadratic (like usual nonrelativistic particles) nor linear (like relativistic particles and Nambu-Goldstone bosons in quantum field theory).
In the unstable configuration (orange and green in Fig.~\ref{fig:unstable_k}A), $\Force < 0$, $\omega$ is real (solid curves) only for wavevectors larger than $k_{\rm c} = \sqrt{-\Force \bar\rho / \sigma}$ (arrows), and becomes imaginary for smaller wavevectors (dashed curves).
The latter describes the RTI where excitations grow exponentially in time.
Here we describe the experimental observation of the RTI as well as ripplon excitations in a ferromagnetically interacting spinor BEC of $^{23}$Na atoms.
We employ matter-wave interference between our two atomic states to transform information regarding the tangential velocity difference across the fluid interface (i.e., counterflow) into an observable pearl-necklace-like chain of vortices.
This interface-velocimetry complements direct measurements of atomic density as the RTI drives rapid growth of fluid counterflow (Fig.~\ref{fig:RT_mushroom}).

Our experiments begin with homogeneous quasi-2D BECs in a square confining potential, which enable the formation of well-defined interfaces at $y=0$ with minimal symmetry breaking (Fig.~\ref{fig:RT_mushroom}).
The BEC's width $W = 114(3)\ \rm \mu m$ greatly exceeds the interface width, given by the spin healing length $\xi_s \approx 2\ \rm \mu m$ (see SM) for $\Na23$ atoms in the $\ket{F=1, m_F=-1} \equiv \ket{\downarrow}$ and $\ket{F=2, m_F=-2} \equiv \ket{\uparrow}$ hyperfine states.
We apply a magnetic field $B_z(x,y)$ along $\ez$ and precisely align the interface in the $\ex$-$\ey$ plane by introducing a small gradient
[$B'\equiv\partial_y B_z = 4.1\ \mu\rm {T/cm}$]
to our otherwise uniform bias field.
This gradient exerts oppositely directed Stern-Gerlach forces with difference $\Force / h = (\mu_\uparrow - \mu_\downarrow) B' / h = 9.3(5) \ \hz_um$ (see SM) on the two spin components due to the differing sign of their magnetic moments $\mu_{\uparrow,\downarrow}$.
Starting with this stable configuration, we induce the RTI by reversing the magnetic field gradient at $t=0$.

Figure~\ref{fig:RT_mushroom} shows the resulting evolution of the RTI at small [$\Force / h = -4.6(3) \, \hz_um$] and large [$\Force / h = -7.7(4) \, \hz_um$] differential force, illustrating the development of characteristic structures over time.
Shortly after entering the unstable configuration, nominally sinusoidal perturbations emerge on the interface and grow with time.
For larger $\Force$, the perturbations appear at larger $k$ (smaller length scale) and grow more rapidly.
Both observations are consistent with the linearized model [Eq. \eqref{eqn:dispersion}], which predicts that unstable modes with wave vector $k$ grow in amplitude with a rate-constant $\Gamma = -{\rm Im}(\omega)$.
Interfacial modes experience gain up to a threshold wavevector $k_{\rm c}\propto\sqrt{\Force}$, with the largest growth rate at $k_{\rm max} = k_{\rm c}/\sqrt{3}$, where $\Gamma$ is maximized (crosses in Fig.~\ref{fig:unstable_k}A).

We quantify the growth of the RTI in terms of the peak-to-peak interface height  $\height_{\rm max}$ and find that $\height_{\rm max}$ grows exponentially in time, as shown in Fig.~\ref{fig:unstable_k}B for $\Force / h = -7.7(4) \, \hz_um$.
The solid curve is an exponential fit to the data with growth rate $\Gamma = 88(3) \ \rm s^{-1}$, consistent with the computed rate $80(5) \ \rm s^{-1}$ (theoretical uncertainty is dominated by that of the magnetic field gradient).
The breakdown of the simple linear model [Eq.~\eqref{eqn:dispersion}] is evidenced by a transition from spatially oscillatory to mushroom- and spike-like interfaces in Fig.~\ref{fig:RT_mushroom}.
At this time, the interface growth slows from exponential to linear before the interface disintegrates, making $\height_{\rm max}$ ill defined.
This overall behavior is reminiscent of RTI dynamics in classical fluids \cite{goncharov2002analytical}.

We obtained the full set of unstable modes by analyzing the power spectral density ${\rm PSD}(k) = |\tilde \height(k)|^2$ of the interface height profile $\height(x)$ [with Fourier transform $\tilde \height(k)$].
The power spectral density (PSD) quantifies the degree of excitation across different spatial frequencies, thereby highlighting modes that were amplified by the RTI.
As shown in Fig.~\ref{fig:unstable_k}C the range of unstable modes increase as the force $\Force$ increases.
For comparison Fig.~\ref{fig:unstable_k}A plots the theoretical dispersion relation for $\Force / h = -3.1 \ \hz_um$ and $-7.7 \ \hz_um$ (corresponding to the small and large $\Force$ presented in Fig.~\ref{fig:RT_mushroom}), computed using Eq.~\eqref{eqn:dispersion}.
In accordance with our experimental observations, the solid curve in Fig.~\ref{fig:unstable_k}C shows the threshold wave vector $k_{\rm c}$, above which modes become stable.

\begin{figure*}[t]
    \centering
    \includegraphics{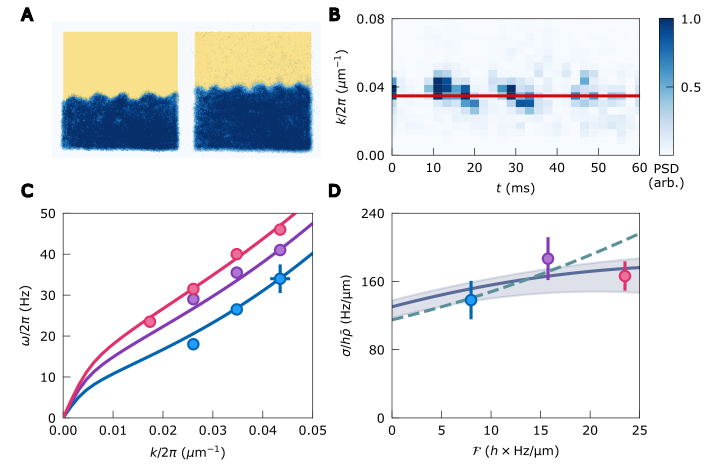}
    \caption{
      \textbf{Excitation and detection of ripplons.}
      (\textbf{A}) Parametrically excited ripplons with $\Force / h = 15.4(8) \ \hz_um$.
      Excitation frequencies $\omega_d / 2\pi = 74\ \Hz$ (left) and  $87\ \Hz$  (right) led to
      $k/2\pi = 0.035 \, \mu\mathrm{m}^{-1}$ and $0.043 \, \mu\mathrm{m}^{-1}$ respectively.
      (\textbf{B}) PSD of the height field of a freely
      evolving interface after excitation. The solid line marks the
      $k$ of the excited mode.
      (\textbf{C}) Ripplon spectrum measured at $\Force / h = 7.7(4) \ \hz_um$ (blue),
      $15.4(8) \ \hz_um$ (purple), and $23(1) \ \hz_um$ (red).
      Dots mark experimental measurements, and curves denote numerical simulations.
      (\textbf{D}) Normalized interface tension.
      The blue solid curve results from fitting Eq.~(\ref{eqn:dispersion}) to the Bogoliubov spectrum.
      The green dashed line is the prediction from the na{\"i}ve model assuming the interface tension stays unchanged across different $\Force$ (see text).
      Error bars and bands mark 1-$\sigma$ uncertainties.
      }
    \label{fig:interface_mode_dispersion}
\end{figure*}

For stable configurations (with $\Force>0$), we employ a parametric driving scheme
to excite ripplon modes with well-defined wavevector $k$ and frequency $\omega$, examples of which are shown in Fig.~\ref{fig:interface_mode_dispersion}A.
After preparing the initial interface with constant $\Force$, we add a small sinusoidal modulation $\delta\Force(t)~\propto\cos(\omega_{\rm d} t)$ to resonantly excite ripplon standing waves.
We note that $\omega_{\rm d}\neq2\omega$ because of the complicated dynamics of the excitation process.
Once the desired mode can be resolved, we remove the sinusoidal drive and allow the system to evolve freely.
Our parametric drive scheme amplifies initial fluctuations with random spatial phase; therefore, we quantify the free evolution in terms of the PSD, a phase-independent quantity.
As seen in Fig.~\ref{fig:interface_mode_dispersion}B, the PSD reveals both the wavevector and frequency of the freely evolving interface waves.

We obtained the ripplon dispersion relation for three values of $\Force$, shown in Fig.~\ref{fig:interface_mode_dispersion}C; the data show good agreement with Bogoliubov spectra (solid curves) computed for our experimental conditions with no adjustable parameters.
As indicated by Eq.~\eqref{eqn:dispersion}, the behavior of these dispersions depend both on the differential acceleration $\Force/m$ (known experimentally) and the ratio of interfacial tension to density $\sigma / \bar\rho$.
Figure~\ref{fig:interface_mode_dispersion}D plots this ratio, obtained by fitting the ripplon dispersion to experimental data (points) or the Bogoliubov spectrum (solid curve).
The inhomogeneous bulk density induced by the magnetic field gradient precludes the direct application of existing theories~\cite{mazets2002waves,barankov2002boundary,vanschaeybroeck2008interface,kobyakov2011rti}, which assume the bulk density is nominally constant over the extent of the interface.
For comparison, we numerically compute $\sigma$ from the difference in free energy with and without the interface present, and obtain $\bar\rho$ using the local density approximation (see SM).
We found nominal agreement between the theory prediction and our observation at small $\Force$, with deviations becoming prominent for larger $\Force$.

\begin{figure*}[t]
    \centering
\includegraphics[width=\textwidth]{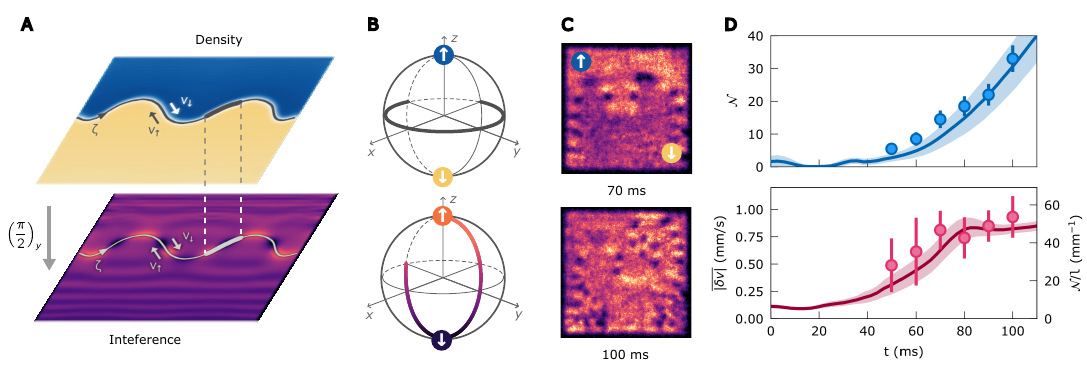}
    \caption{
    \textbf{Interface velocimetry.}
    (\textbf{A}) Interferometry scheme showing the original density distributions $\rho_{\uparrow,\downarrow}$ (top) and a simulated interference pattern in $\rho_{\uparrow}$ after the $\pi/2$-pulse (bottom).
    The solid curve marks the interface; $\zeta$ is the spatial coordinate defined on the interface;
    and the arrows indicates the fluid velocity on each side of the interface.
    (\textbf{B}) Spin vector $\Psi$ on the Bloch sphere before (top) and after (bottom) the $\pi/2$-pulse.
    The arcs colored in grey (top) and orange/purple (bottom) identify the spin-orientation on the bold segments in A.
    (\textbf{C}) Experimental images of $\rho_\uparrow$ after the microwave $\pi/2$-pulse.
    (\textbf{D}) Time evolution of the vortex number (top) and average magnitude of the interfacial counterflow velocity $\overline{| \delta v|}$ (bottom).
    The markers represent measurements, and the solid curves are numerical results.
    Error bars and bands mark 1-$\sigma$ uncertainties.
    }
    \label{fig:interface_velocimetry}
\end{figure*}

We conclude by using the binary BEC's phase degrees of freedom $\phi_{\uparrow,\downarrow}(\mathbf r)$ to study the superfluid velocity at the interface.
Each component of the BEC is described by a macroscopic wavefunction, for example $\psi_\uparrow(\mathbf r) = \sqrt{\rho_\uparrow(\mathbf r)} \exp[i\phi_\uparrow(\mathbf r)]$ for the $\ket{\uparrow}$ component, where the fluid velocity ${\mathbf v}_\uparrow = \hbar {\boldsymbol \nabla} \phi_\uparrow / m$ derives from spatial phase gradients.
As a result, the phase difference $\delta \phi = \phi_\uparrow - \phi_\downarrow$ at the interface encodes a velocity difference $\delta v = \hbar {\partial_\zeta} \delta \phi / m$.
This describes interfacial counterflow (indicated in Fig.~\ref{fig:interface_velocimetry}A), as a function of position $\zeta$ on the interface.
Our two-component BEC is described by a combined spinor wavefunction $\Psi(\mathbf r) = [\psi_{\uparrow}(\mathbf r), \psi_{\downarrow}(\mathbf r)]^T$, visualized at each point in space by a vector on the Bloch sphere (Fig.~\ref{fig:interface_velocimetry}B).
For example, in the $\ket{\uparrow}$ domain this vector points to the north-pole, and conversely to the south-pole in the $\ket{\downarrow}$ domain; at the interface, where the spin components overlap with equal density, the Bloch vector lies in the equatorial plane with azimuthal angle equal to $\delta \phi$.

Experimentally, a microwave $\pi/2$-pulse (driving rotations about the $y$-axis of the Bloch sphere) interferes the two macroscopic wavefunctions; extended regions of interfacial counterflow (associated with helical spin textures prior to the $\pi/2$-pulse) are transformed into interpenetrating vortex-chains in each component with zero-density vortex cores in one component aligned with density peaks in the other.
As illustrated in (Fig.~\ref{fig:interface_velocimetry}A), this vortex-chain follows the interface, where the densities of the individual components were initially equal, and is a sinusoidal function
$\rho_\uparrow = \bar\rho \cos^2 (\delta \phi/2)$ of the initial phase difference.

Figure~\ref{fig:interface_velocimetry}C shows experimental interference patterns as a function of time $t$ after initiating the RTI [with $\Force / h =4.6(3)\ \hz_um$].
The total number of vortices $\mathcal{N}$ indicates how many times $\delta \phi$ changed by $2\pi$ at the interface (of length $L$); this can be readily linked to the average magnitude of the counterflow velocity through
$\overline{|\delta v|} \approx h\mathcal{N}/(m L)$ with the uncertainty attributed to the limited vortex density and regions where $\delta v$ switches its direction.
Figure~\ref{fig:interface_velocimetry}C shows the result of this analysis:
at short times (linear dynamics), the increasing value of $\mathcal{N}$, with zeros following well-defined interface mark the initial development of counter-currents.
At longer times (non-linear and turbulent dynamics), the growth of $\overline{|\delta v|}$ stops, while increasingly complex interfacial patterns lead to a proliferation of vortices that expose the intricate pattern of currents associated with domain formation.

Our observation of the RTI together with the excellent agreement with theory expands the repertoire of fluid dynamical instabilities observed in ultracold atom systems.
Recent work established that superfluid counterflow drives the Kelvin-Helmholtz instability, leading to the development of conventional vortex chains in single-component systems~\cite{hernandez2024connecting,mukherjee2022quantumhall} and more exotic spin textures in multi-component systems~\cite{Huh2024}.
Together, these works highlight the utility of ultracold atom experiments for realizing (super-) fluid instabilities in settings with well calibrated microscopic parameters and without the full complexity of the usual Naiver-Stokes equations.

Highly elongated and precisely aligned interfaces between quantum fluids, such as those generated here, offer new technical and scientific opportunities.
From a technical standpoint, the equilibrium distribution of ripplon modes---amplified by the RTI and measured using PSDs such as presented here---have immediate applicability to thermometry of BECs.
In particular, the $\omega\propto k^{3/2}$ dispersion as $\Force \rightarrow 0$ allows access to ultra-low energy thermal excitations.
From a scientific perspective, superfluid interfaces provide a new platform for analog simulation of relativistic field theories in the post-inflationary thermalization in the early universe~\cite{fifer2019analog,barroso2023non}.

Our technique for parametrically exciting the ripplon modes can be interpreted as a (Floquet) process that destabilizes specific otherwise stable interface modes.
Reference ~\cite{kobyakov2012parametric} proposed the complementary question for binary BECs:
can parametric modulation stabilize the superfluid RTI?
Answering this question has important ramifications for a host of fluid dynamics problems where RTI is detrimental.
For example, modulating the accelerating force is expected to suppress RTI effects that destabilize inertial confinement fusion~\cite{Boris1977}
and has been observed to stabilize fluid-gas interfaces~\cite{Apffel2020}.

{\it Author Contributions}---
Y.G. and J.T conducted the experiment (initially as a secret project);
Y.G., J.T, and M.Z. performed theoretical work;
Y.G analyzed the data;
Y.G. and S.M. developed, implemented, and maintained the experimental apparatus.
Y.G., J.T., and M.Z. conceived of the experiment with input from I.B.S.;
S.E., G.K.C. and I.B.S. provided mentorship;
and G.K.C. and I.B.S. obtained funding.
All authors substantially participated in the
discussion and the writing of the manuscript.

{\it Acknowledgments}---
The authors thank N.~H.~Pilgram and D.~Kurdak for carefully reading the manuscript.
This work was partially supported by the National Institute of Standards and Technology; the National Science Foundation through the Quantum Leap Challenge Institute for Robust Quantum Simulation (grant OMA-2120757); and the Air Force Office of Scientific Research Multidisciplinary University Research Initiative ``RAPSYDY in Q'' (FA9550-22-1-0339).

\bibliography{main}

\newpage
\cleardoublepage

\setcounter{figure}{0}
\setcounter{equation}{0}

\onecolumngrid

\section{Supplementary Information}

\subsection{Initial state preparation}

We prepare quasi-2D $\Na23$ BECs with $N\approx 1\times 10^6$ atoms and 2D chemical potential $\chemicalpotential / h \approx 600 \ \rm Hz$ in the $\ket{F=1, m_F=-1} \equiv \ket{\downarrow}$ ground state in a uniform magnetic field $B_z = 70 ~\mu\rm T$, vertically aligned along $\ez$.
The BECs are tightly confined along $\ez$ by a harmonic trap with frequency $\omega_z/ (2\pi)= 1.1~\rm kHz$.
At the same time, they are horizontally enclosed in a square potential with width $W = 114(3)\ \rm \mu m$, creating a homogeneous density profile in the $\ex-\ey$ plane.
To create the spinor interface, we apply a microwave $\pi/2$ pulse to transfer half of the population to $\ket{F=2, m_F=-2}\equiv\ket{\uparrow}$.
Because these two spin states are immiscible [scattering lengths $a_{\uparrow\uparrow} = 62.5(5) a_B, a_{\downarrow\downarrow} = 54.54(20) a_B$,
$a_{\uparrow\downarrow} = 64.3 a_B$
see Ref. \cite{Samuelis2000,Knoop2011}
and footnote\footnote{This number is first reported in \cite{lamporesi2023twocomponentspinmixtures}.
We assumed it has an uncertainty of $0.4 a_B$, similar to other reported
scattering length in \cite{Knoop2011}.}
; giving $a_{\uparrow\uparrow} a_{\downarrow\downarrow}< a_{\uparrow\downarrow}^2$], they spontaneously phase separate.
We align the interface of the final phase separated mixture by applying a magnetic field gradient $B'=- 4.1~\mu \rm T/cm$ for approximately 2~s.

\subsection{Imaging}

We measure the 2D atomic density $\rho_\uparrow$, i.e. density of $\ket{F=2, m_F=-2}$ atoms, using resonant absorption imaging.
We illuminate the atomic ensemble with a probe laser beam traveling along $\mathbf{e}_z$ resonant with the $\ket{F=2, m_F=-2} \to \ket{F=3, m_F=-3}$ cycling transition.
The resulting optical field is collected by an objective lens with numerical aperture ${\rm NA} \approx 0.3$ and imaged on a charge coupled device (CCD) camera.

Given the limitations of our camera, we are unable to image  $\rho_\downarrow$ in the same experimental run.
Instead, we infer $\rho_\downarrow$ by subtracting a binarized image of $\rho_\uparrow$ from that of a homogeneous condensate in the same potential; this yields the yellow regions presented in Figs.~1 and 3 in the main text.
The data presented in Fig.~4 was measured in the same way, but the interferometry procedure naturally yielded a nominally uniform density in  $\rho_\uparrow$.

\subsection{Interface dynamics in a classical fluid}

Here we briefly derive the gravity wave dispersion relation of a pair of immiscible classical fluids.
We begin with Euler's equation in potential form
\[
\frac{\partial \phi }{\partial t}+\frac{p}{\rho }+\frac{F}{m}y=0,
\]
describing the flow of an incompressable fluid with particle mass $m$, density $\rho$, velocity potential $\phi$ and pressure $p$, subject to an applied force $F$.
The Young-Laplace equation
\cite[\S 61]{landau1987fluid}
\[
\Delta p=-\sigma \frac{\partial ^{2}\eta }{\partial x^{2}},
\]
describes the effect of interface tension resulting from a 1D interface nominally aligned along $\ex$, where $\Delta p$ is the pressure difference across the interface and $\eta$ is the vertical displacement of the interface.
This problem can be solved by imposing the boundary condition that velocity of both fluids normal to the interface are be equal, in which case one obtains wave-like solutions $\eta \propto \exp\left[i(kx-\omega t)\right]$ with wavenumber $k$, angular frequency $\omega$, and dispersion relation
\cite[\S 62]{landau1987fluid}
given as  Eq.~(1) in the text
\begin{align}
    \omega ^{2}&=\frac{1}{2m} \left(\mathcal{F} k+\frac{\sigma}{\bar \rho} k^{3}\right),
\end{align}
with density-weighted force difference
\begin{align}
    \mathcal{F} &=  B' \left(\mu_\uparrow \frac{\rho_{\uparrow}}{ \bar \rho} -  \mu_\downarrow \frac{\rho_{\downarrow}}{ \bar \rho}\right).
\end{align}
Unlike for our BECs, in incompressible fluids the individual component $\rho_{\uparrow,\downarrow}$, and averaged $\bar{\rho}=(\rho_\uparrow+\rho_\downarrow)/2$ bulk densities are homogeneous and therefore unambiguous.

The force difference can be cast in the form of a per particle differential force $\mathcal{F} =  B' (\mu^*_\uparrow - \mu^*_\downarrow)$ by introducing renormalized magnetic moments
\begin{align}
\mu^*_{\uparrow\downarrow} &= \mu_{\uparrow\downarrow} \left(1 \pm \frac{\delta\rho}{\bar\rho}\right).
\end{align}
Thus, $\mathcal{F}$ reduces to the true per particle differential force $B'(\mu_\uparrow  -  \mu_\downarrow)$ when the volume-densities are equal ($\delta\rho \equiv \rho_\uparrow - \rho_\downarrow = 0$), as well as when the magnetic moments are equal and opposite ($\mu_\uparrow  = - \mu_\downarrow$).
In our experiment, $\delta\rho \neq 0$ results from the $14~\%$ fractional difference between the scattering lengths $a_{\uparrow\uparrow}$ and $a_{\downarrow\downarrow}$; the resulting correction to $\mu_{\uparrow\downarrow}$ leads to an $\approx 4~\%$ change in $\mathcal F$, comparable the uncertainties from other sources.
Because we directly compare to the results of Bogoliubov-de Gennes and  Gross-Pitaevskii calculations, this only impacts the determination  of $\sigma / \bar\rho$ in Fig.~3.

\subsection{Governing equations for the dynamics of a binary condensate}

The Hamiltonian describing a weakly interacting binary bosonic field is
\begin{align*}
\hat H =&  \int {\rm d}^3 {\bf r} \left\{
\hat \psi_{\uparrow}^{\dagger}({\bf r}) \left[ -\frac{\hbar ^{2}}{2 m} \nabla^2 -\chemicalpotential _\uparrow+V_\uparrow\left( {\bf r}\right) \right] \hat \psi_{\uparrow}({\bf r})  + \hat \psi _\downarrow^{\dagger}({\bf r}) \left[ -\frac{\hbar ^{2}}{2 m} \nabla^2-\chemicalpotential _{\downarrow}+V_\downarrow\left( {\bf r}\right) \right] \hat \psi_{\downarrow}({\bf r}) \right. \\
 &\left. +\frac{g_{\uparrow\uparrow}}{2}\hat \psi _{\uparrow}^{\dagger}({\bf r})\hat \psi _{\uparrow}^{\dagger}({\bf r}) \hat \psi_{\uparrow}({\bf r}) \hat \psi_{\uparrow}({\bf r}) +\frac{g_{\downarrow\downarrow}}{2}\hat \psi _{\downarrow}^{\dagger}({\bf r})\hat \psi_{\downarrow}^{\dagger}({\bf r}) \hat \psi_{\downarrow}({\bf r}) \hat \psi_{\downarrow}({\bf r}) + g_{\uparrow\downarrow}\hat \psi _{\uparrow}^{\dagger}({\bf r})\hat \psi _{\downarrow}^{\dagger}({\bf r}) \hat \psi_{\downarrow}({\bf r}) \hat \psi_{\uparrow}({\bf r}) \right\},
 \end{align*}
where $\hat \psi_{\uparrow,\downarrow}$ are the bosonic field operators; $\chemicalpotential_{\uparrow,\downarrow}$ are the state-dependent chemical potentials; $g_{\uparrow\uparrow}$ and $g_{\downarrow\downarrow}$ are the intra-species interaction strengths; and $g_{\uparrow\downarrow}$ is the inter-species interaction strength.
The potentials $V_{\uparrow,\downarrow}({\bf r})$ result from the combined state-independent optical box potential and the spin-dependent magnetic gradient potential.

The Heisenberg picture equations of motion for the field operators
\[
i \hbar \frac{\partial}{\partial t} \hat \psi_{\uparrow,\downarrow} ({\bf r}, t) =
[\hat \psi_{\uparrow,\downarrow} ({\bf r}, t), \hat H],
\]
lead to the the coupled equations
\begin{align}
i \hbar \frac{\partial}{\partial t} \hat \psi_\uparrow &= \left[-\frac{\hbar ^{2}}{2m} \nabla^2 - \chemicalpotential_\uparrow
+ V_\uparrow({\bf r})\right] \hat \psi_\uparrow
+ g_{\uparrow\uparrow} \hat \psi_\uparrow^\dagger \hat \psi_\uparrow \hat \psi_\uparrow
+ g_{\uparrow\downarrow} \hat \psi_\downarrow^\dagger \hat \psi_\downarrow \hat \psi_\uparrow\\
 i \hbar \frac{\partial}{\partial t} \hat \psi_\downarrow &= \left[-\frac{\hbar ^{2}}{2m} \nabla^2 - \chemicalpotential_\downarrow
+ V_\downarrow({\bf r})\right] \hat \psi_\downarrow
+ g_{\downarrow\downarrow} \hat \psi_\downarrow^\dagger \hat \psi_\downarrow \hat \psi_\downarrow
+ g_{\uparrow\downarrow} \hat \psi_\uparrow^\dagger \hat \psi_\uparrow \hat \psi_\downarrow.
\end{align}
Assuming that a single mode is macroscopically occupied, as for a Bose-Einstein condensate (BEC), we replace the field operators with scalar fields $\hat \psi_{i}({\bf r}, t) \to \psi_{i}({\bf r}, t)$.
The resulting equations

\begin{align}
i \hbar \frac{\partial}{\partial t} \psi_\uparrow &= \left[-\frac{\hbar ^{2}}{2m} \nabla^2 - \chemicalpotential_\uparrow
+ V_\uparrow({\bf r})\right] \psi_\uparrow
+ g_{\uparrow\uparrow} |\psi_\uparrow|^2 \psi_\uparrow
+ g_{\uparrow\downarrow} |\psi_\downarrow|^2 \psi_\uparrow \label{eqn:gpe1}\\
 i \hbar \frac{\partial}{\partial t} \psi_\downarrow &= \left[-\frac{\hbar ^{2}}{2m} \nabla^2 - \chemicalpotential_\downarrow
+ V_\downarrow({\bf r})\right] \psi_\downarrow
+ g_{\downarrow\downarrow} |\psi_\downarrow|^2 \psi_\downarrow
+ g_{\uparrow\downarrow} |\psi_\uparrow|^2 \psi_\downarrow \label{eqn:gpe2}
\end{align}
are spinor Gross-Pitaevskii equations (GPEs), and we denote the spinor ground state as $\Psi^{(0)}({\bf r}) = \left[\psi^{(0)}_\uparrow({\bf r}), \psi^{(0)}_\downarrow({\bf r})\right]^T$.

Several length scales are associated with Eq.~\eqref{eqn:gpe1} and \eqref{eqn:gpe2}.
First, in spin-polarized domains the healing lengths  $\xi_{\uparrow,\downarrow} = \sqrt{\hbar^2/(2m\chemicalpotential_{\uparrow,\downarrow})}$ describe the spatial scale associated with density perturbations; since $\chemicalpotential_\uparrow \approx \chemicalpotential_\downarrow$ in equilibrium we have $\xi_{\uparrow} \approx \xi_{\downarrow}$ with  $\xi \equiv (\xi_{\uparrow} + \xi_{\downarrow}) / 2 $.
Similarly, the spin healing length $\xi_s/\xi = \sqrt{\bar g / (g_{\uparrow\downarrow} - \bar g)}$ describes the spatial extent of the interface between domains~\cite{ao1998segregate}, where $\bar g = \sqrt{g_{\uparrow\uparrow}g_{\downarrow\downarrow}}$ is an average interaction strength.
In our experiment, $\xi \approx 0.6 \ \rm{\mu m}$ and $\xi_s \approx 2 \ \rm{\mu m}$.

\subsection{Excitations}

Small excitations above the condensate ground state can be analyzed by splitting the GPE wavefunction into $\Psi({\bf r}) =
\Psi^{(0)}({\bf r}) + \delta \Psi({\bf r})$, where $\delta \Psi({\bf r})$ describes small fluctuations of the GPE wavefunction.
At linear order, this leads to a set of equations known as the Bogoliubov-de Gennes (BdG) equations~\cite{Takahashi2015}.

We simulated the dynamics of the system with a GPU-based 2D GPE solver, implemented using a finite-difference method.
The use of a 2D solver is valid owing to the quasi-2D nature of the BEC: the transverse confining frequency  $\omega_z/2\pi \approx 1.1 \  \rm kHz$ is larger than the $\chemicalpotential_{\uparrow,\downarrow}/h \approx 0.7\ {\rm kHz}$ chemical potential.
The simulation parameters are carefully selected to match the experimental conditions, with effective 2D interaction constants
$g^{(2D)}_{ij} = g_{ij}\sqrt{m \omega_z/(2\pi\hbar)}$~\cite{Dalibard2011}, with $i,j \in \{\uparrow, \downarrow\}$.
We calibrate the effective 2D chemical potential by matching the frequency of the fundamental sound mode \cite{ville2018firstsound} with the experimental value of $14.9(3)\ {\rm Hz}$ for a spin polarized sample in $\ket{F=1, m_F=-1}$.

To model ripplon excitations in the stable configuration, we first compute the ground state by evolving in imaginary time~\cite{Lehtovaara2007imaginarytime}.
Rather than explicitly solving the 2D BdG problem, we consider longitudinal traveling wave solutions of the form $\delta\Psi(y)e^{ikx}$ with wavenumber $k$ \cite{sasaki2009rti,takahashi2015ngmodes}, and use Arnoldi iteration to obtain the lowest several BdG eigenvectors and eigenvalues of the resulting 1D problem.

\begin{figure}[t]
    \centering
    \includegraphics{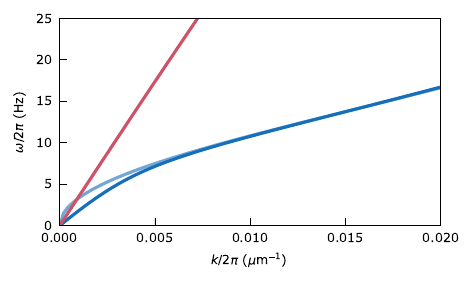}
    \caption{
    Dispersion of the two lowest gapless modes.
    The dark blue and red curve show the dispersion relations of the ripplon and phonon modes respectively.
    The light blue curve is a fit of the ripplon dispersion to Eq.~(1), from which we could obtain $\sigma /\bar\rho$.
}
    \label{fig:two_modes}
\end{figure}

Figure~\ref{fig:two_modes} shows energies of the lowest two modes with the ripplon mode in dark blue, and the phonon mode in red; in addition the light blue curve plots the ripplon dispersion from Eq.\ (1).
At small $k$ the phonon mode crosses the idealized ripplon dispersion, leading to an avoided crossing like behavior and a breakdown of the $\omega\propto\sqrt{k}$ scaling at very small $k$.

\subsection{Extracting $\sigma/\bar\rho$}

The interfacial tension to density ratio $\sigma / \bar\rho$ plotted in Fig.~3(D) was obtained from a least squares fit of Eq.~(1) to each experimentally measured or numerically computed dispersion; for example, the light blue curve in Fig.~\ref{fig:two_modes} is a fit to the dark blue curve.
For large wavenumber, i.e. $k \xi_s \gtrsim1$, the internal spatial structure of the domain wall becomes relevant, where our simulations yield  $\omega\propto k^{5/2}$ behavior.
At yet larger $k$, when $k \xi \gtrsim1$, excitations recover their free-particle structure leading to a conventional $\omega\propto k^2$ dispersion.
As a result, choosing different upper limits of $k$ when performing the fit to the BdG dispersion leads to a slightly different values of $\sigma/\rho$.
In Fig. 3(D), the error band around the theory curve is produced by varying this upper limit from $0.025 \ \rm {\mu m}^{-1}$ to $0.08 \ \rm {\mu m}^{-1}$ (corresponding to $k \xi_s$ ranging from $0.3$ to $1$).

\begin{figure}[t]
    \centering
    \includegraphics{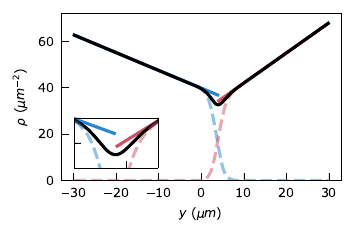}
    \caption{Spin-resolved density profile under $\Force / h =
7.3(4) \ \hz_um$ with $\uparrow$/$\downarrow$ and their sum in blue/red and black respectively.
The numerically computed profiles (dashed blue and red)
are overlaid with corresponding LDA estimates (solid blue and red) that ignore
the inter-spin interaction at the interface. The inset shows an expanded view
of the region near interface.
}
    \label{fig:density}
\end{figure}

Figure~3(D) also compares our result to analytical theories; the approximations introduced in Refs.~\cite{kobyakov2011rti,vanschaeybroeck2008interface,mazets2002waves} predict
\begin{align}
    \sigma = 2 \int_{-\infty}^{+\infty} {\rm d} y \ \frac{\hbar^2}{2m}\left( |\partial_y \psi_\uparrow|^2 + |\partial_y \psi_\downarrow|^2 \right) \label{eqn:sigma}
\end{align}
for condensates with homogeneous bulk densities.
Because atomic BECs are compressible, the magnetic gradient leads to non-vanishing derivatives $\partial_y \psi_{\uparrow\downarrow}$ even far from the interface, as suggested by Fig.~\ref{fig:density}.
This impedes a straightforward application of Eq.~\eqref{eqn:sigma}: the interfacial tension $\sigma$, defined as the excess free energy due to the interface, should exclude the explicit contribution to the energy from the magnetic gradient potential.

Existing results~\cite{vanschaeybroeck2008interface,Takeuchi2010} account for inhomogeneous potentials by applying Eq.~\eqref{eqn:sigma} to homogeneously trapped systems with atom density at the interface equal to that of a corresponding inhomogeneous case, i.e., a form of local density approximation (LDA).
This approach breaks down when the individual component densities change significantly on the spatial scale of the interface, as for our strong magnetic gradient potentials; this leads to an underestimate of the interfacial tension.
To circumvent this difficulty, we directly compute the per-unit length difference in free energies
\begin{align}
\sigma = \left[\Omega(\psi_\uparrow, \psi_\downarrow) - \Omega(\psi_{\text{LDA}, \uparrow}, \psi_{\text{LDA}, \downarrow})\right]/L \label{eqn:free_energy_diff},
\end{align}
between that of our experimental conditions (with mode functions $\psi_{\uparrow,\downarrow}$) and that predicted by the LDA [with $\psi_{\text{LDA}, j} = \sqrt{(\epsilon_j - V_j)/g_{jj}}]$; the densities for both cases densities are plotted in Fig.~\ref{fig:density}).
Here the free energy is
\begin{align*}
\Omega(\phi_\uparrow, \phi_\downarrow) =&  \int {\rm d} x\ {\rm d} y \left\{
\frac{\hbar ^{2}}{2 m} | \nabla \phi_{\uparrow} |^2 + \frac{\hbar ^{2}}{2 m} | \nabla \phi_{\downarrow} |^2 +
\left[
V_\uparrow\left( y\right) -\chemicalpotential_\uparrow \right]|\phi_{\uparrow}(y)|^2
+ \left[V_\downarrow\left( y\right)  -\chemicalpotential_{\downarrow} \right] |\phi_{\downarrow}(y)|^2 \right. \\
 &\left. +\frac{g_{\uparrow\uparrow}}{2}|\phi_{\uparrow}(y)|^4 +\frac{g_{\downarrow\downarrow}}{2}|\phi _{\downarrow}(y)|^4 + g_{\uparrow\downarrow}|\phi_{\uparrow}(y)|^2 |\phi_{\downarrow}(y)|^2 \right\}.
\end{align*}
The resulting ratio $\sigma / \bar\rho$ is plotted as the green dashed curve in Fig.~3D.

As the inhomogeneous potential approaches zero, the individual contributions to Eq.~\eqref{eqn:free_energy_diff} return to those of the homogeneous system, with free energy $\Omega(\psi_{\text{LDA}, \uparrow}, \psi_{\text{LDA}, \downarrow}) = -PV$, pressure $P = \chemicalpotential_\uparrow^2 / 2 g_{\uparrow\uparrow} = \chemicalpotential_\downarrow^2 / 2 g_{\downarrow\downarrow}$, and volume $V$.
Under these conditions~\cite{vanschaeybroeck2008interface}, Eq.~\eqref{eqn:free_energy_diff} reduces to Eq.~\eqref{eqn:sigma}.

\subsection{Unstable ripplon modes}

Similar to our approach the stable configuration, we begin by using imaginary time evolution to obtain the metastable ``ground'' state under the assumption of translational invariance along $\ex$.
In this configuration, the tendency of the two spin components to evolve to the real ground state can be stymied by the ferromagnetic interaction $g_{\uparrow \downarrow}$.
We then continue as above and perform a BdG analysis about this metastable state yielding dispersions as shown in Fig.~2(C); the interfacial tension ratio $\sigma/\bar\rho$ is then  obtained from the critical wavevector $k_c$.
For sufficiently large gradients the metastable state is no longer stationary under imaginary time evolution.

We partly mitigate this introducing a localized gradient potential that is selected to match the experimental gradient near $y=0$ that then becomes constant for large $|y|$.
This is justified because, for ripplon modes, $\delta\Psi$ is only significant near the interface.
However, for $B'>5.3\ \hz_um$ the tendency to find the true ground state cannot be avoided.

\begin{figure}[t]
    \centering
    \includegraphics{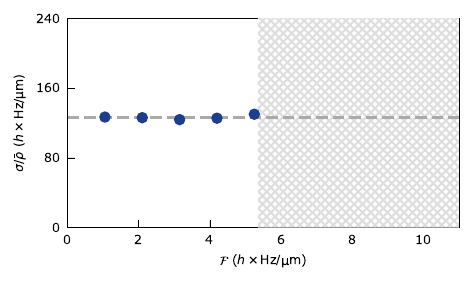}
    \caption{
    Interfacial tension to density ratio.
    Symbols show $\sigma/\bar \rho$ acquired from the BdG spectrum, with average given by the dashed line.
    The hatched region marks parameters for which we were unable to relax to the metastable state using imaginary time evolution.
}
    \label{fig:sigma_rho}
\end{figure}

Figure \ref{fig:sigma_rho} plots the numerically obtained ratio $\sigma/\bar\rho$ up to this threshold, showing that it is largely independent of $\Force$; the range of the horizontal axis matches those explored experimentally.
This shows that $\sigma/\bar\rho$ is nearly independent of $\Delta\Force$.
To construct the theory curves in Fig. 2(C), we simply take the average value of $\sigma/\bar\rho$ (dashed line in Fig.~\ref{fig:sigma_rho}) to extrapolate into the parameter regime where our numerics break down.

\subsection{Interface velocimetry}

The initial state is initialized with actual experimental parameters.
We model the system's finite temperature by introducing random phase noise into the GPE ground state and evolving forward in time until steady state is achieved.
The amount of injected noise is determined by matching the overall amplitude (not rate) of the exponentially-growing peak-to-peak interface height $\eta_{\rm max}$ to the experimental observations in Fig.~2.
The length of the interface $L$ is acquired from the simulations.
The vortex number $\mathcal N$ is extracted manually for experimental data and with a counting algorithm for simulations.

\end{document}